\documentclass[aps,epsfig]{revtex4}
\usepackage{graphicx}
\begin{document}
\def\be{\begin{equation}}
\def\ee{\end{equation}}
\def\bfi{\begin{figure}}
\def\efi{\end{figure}}
\def\bea{\begin{eqnarray}}
\def\eea{\end{eqnarray}}

\title{Fluctuations and effective temperatures in coarsening}

\author{Federico Corberi}
\affiliation {Dipartimento di Matematica ed Informatica, Universit\`a di Salerno, 
via Ponte don Melillo, 84084 Fisciano (SA), Italy.}
\affiliation{Universit\'e Pierre et Marie Curie - Paris VI\\
Laboratoire de Physique Th\'eorique et Hautes Energies\\
4 Place Jussieu, 5\`eme \'etage, 75252 Paris Cedex 05, France}
\author{Leticia F. Cugliandolo}
\affiliation{Universit\'e Pierre et Marie Curie - Paris VI\\
Laboratoire de Physique Th\'eorique et Hautes Energies\\
4 Place Jussieu, 5\`eme \'etage, 75252 Paris Cedex 05, France}

\begin{abstract}
  We study dynamic fluctuations in non-disordered finite dimensional
  ferromagnetic systems quenched to the critical point and the
  low-temperature phase.  We investigate the fluctuations of two
  two-time quantities, called $\chi$ and $C$, the averages of
  which 
  yield the self linear response and correlation function. We
  introduce a restricted average of the $\chi$'s, summing over all
  configurations with a given value of $C$.  We find that
  the restricted average $\langle \chi
  \rangle_C$ obeys a scaling form, and that the slope of the scaling
  function approaches the universal value $X_\infty $ of the limiting
  effective temperature in the long-time limit and for $C\to 0$.  Our
  results tend to confirm the expectation that time-reparametrization
  invariance is not realized in coarsening systems at
  criticality. Finally, we discuss possible experimental tests of our
  proposal.
\end{abstract}

\maketitle


\section{Introduction} \label{intro}

Fluctuation-dissipation relations (FDRs), namely
model-independent relations between linear response functions and correlation functions,
have been extensively investigated 
in systems that relax slowly out of 
equilibrium~\cite{Cuku,Corberi-review,Caga,Godreche-review,Crri,Cukupe,fmpp}. 
Special emphasis was set on the 
analysis of aging cases. For spin systems
the impulsive auto-response function, describing the effect of a
perturbing magnetic field $h_i(t')$ acting on site $i$ at time $t'$ on
the magnetization $\langle s_i(t)\rangle$ on the same site $i$ at the
later time $t>t'$, is
\be
\langle R_i\rangle (t,t')=\lim _{h_i\to 0}
\frac{\delta\langle s_i (t)\rangle}{\delta h_i(t')}
\label{r}
\; . 
\ee 
The (spatially averaged) integrated auto-response function, or dynamic susceptibility, is
\be
\langle \chi\rangle (t,t_w)= 
\frac{1}{N}\sum _i\int _{t_w}^t \langle R_i \rangle(t,t') dt'
\; ,
\label{chi}
\ee
where $N$ is the number of spins in the system. For 
ferromagnetic coarsening dynamics it is convenient to consider a
perturbation that is not correlated with the equilibrium ordered
states and quenched random fields $h_i$ are typically used. For
instance, in the case in which 
a bimodal random field, $h_i=h\epsilon_i$, with $h$
the amplitude and $\epsilon_i =\pm 1$ with probability a half is applied,
the susceptibility~(\ref{chi}) can be cast as
\be 
\langle \chi\rangle (t,t_w)= 
\lim _{h\to 0}\frac{1}{N}\sum _i\frac{\langle (s_i^h(t)-s_i(t)) \epsilon_i\rangle}{h} \; ,
\label{eq:discrete-r}
\ee
where $s_i^h(t)$ is the value that the $i$-th spin takes at time $t$ in a trajectory in which the perturbation was switched on from $t_w$ onwards,
and $s_i(t)$ is the value of the spin in a freely 
evolving trajectory. Other choices of random perturbations with, for instance, finite 
spatial correlation are also of interest~\cite{Sollich-fields} but we do not 
use them here.
The average $\langle \dots\rangle$ 
is taken over all possible initial conditions,
thermal histories, the random field  and quenched disorder (if present). 
With this notation, $\langle \chi \rangle$ and $\langle R\rangle$,  
we make explicit the fact that we average over all sources of fluctuations. 

In equilibrium, $\langle \chi\rangle $ and the auto-correlation function
\be
\langle C\rangle(t,t_w)= \frac{1}{N}\sum _i\langle s_i(t) s_i(t_w)\rangle
-\frac{1}{N}\sum _i\langle s_i(t)\rangle \frac{1}{N}\sum _i\langle s_i(t_w)\rangle
\; , 
\label{autocorre}
\ee
computed with the same complete average,
depend only upon the difference $t-t_w$, due to stationarity,
and they are related
through the fluctuation dissipation theorem (FDT):
$T\langle \chi\rangle(t-t_w)= \langle C\rangle(t,t)-\langle C\rangle(t-t_w)=1-\langle C\rangle(t-t_w)$
(we consider here and in the following unitary modulus spins).
In generic non-equilibrium states,
$\langle C\rangle$ is no longer stationary but it is, usually,  a monotonically
decaying function of $t$. One can then invert the relation
between $t$ and $\langle C\rangle$ to obtain
\be
\langle \chi\rangle (t,t_w)=\widehat  \chi (\langle C\rangle,t_w)
\; .
\label{eq:average-chi-C}
\ee
Also quite generally $\widehat  \chi (\langle C\rangle,t_w)$ looses the $t_w$ dependence
at large $t_w$ and the limiting form
\be
\widehat  \chi (\langle C\rangle)=\lim _{t_w\to \infty} \widehat  \chi (\langle C\rangle,t_w) 
\label{eq:limit_chi}
\ee
is a non-trivial function of $\langle C\rangle$ \cite{Cuku,fmpp}.
Moreover, the slope $X(\langle C\rangle)=-Td\widehat \chi (\langle C\rangle)/d\langle C\rangle$ 
allows for the definition of an effective temperature~\cite{Cukupe} through
$T_{eff}(\langle C\rangle)=T/X(\langle C\rangle)$.  For coarsening systems
quenched to the critical point 
the limiting value
\be
X_\infty=-T \, 
\lim _{t_w\to \infty}\lim_{\langle C\rangle\to 0} \frac{d\widehat \chi (\langle C\rangle,t_w)}{d\langle C\rangle} 
\; 
\label{xinfty}
\ee
is of particular relevance~\cite{Godreche-Luck,Corberi-review,Godreche-review,Caga,Sollich-Xinfty,Caga-Xinfty,Corberi-Xinfty,Sollich-fields}. 
Note that taking $\langle C\rangle \to 0$ first implies that one takes $t\to\infty$ before $t_w\to \infty$. 

What discussed insofar shows that important properties of the system,
the effective temperature in particular, are encoded in the dependence
of $\langle \chi\rangle $ on $\langle C\rangle$, with $t$ and $t_w$ being parameters the variation 
of which merely allows one to scan sectors with different values of
the relevant quantity $\langle C\rangle$.  In this paper, we elaborate on this
idea by studying the {\it fluctuations} of these two-time functions.
In complete generality, for a given initial condition, thermal noise 
and random field realization we consider the fluctuating quantities
\begin{eqnarray}
C(t,t_w) &=& \frac{1}{N} \sum_{i=1}^N s_i(t) s_i(t_w) 
\label{eq:C}
\; , 
\\
\chi(t,t_w) &=& 
\lim_{h\to 0}  \frac{1}{N} \sum_{i=1}^N \frac{(s^h_i(t)- s_i(t)) \epsilon_i}{h} 
\; ,
\end{eqnarray}
without any averaging (we dropped the second term on the right hand side of 
Eq.(\ref {autocorre}) since it is negligible for large $N$ in the 
cases with $\langle s_i\rangle \equiv 0$ considered in the following). 
The joint probability distribution $P(C,\chi)$  has
been studied in disordered spin models~\cite{TRI-SG,Jaubert}, some kinetically constrained spin 
systems~\cite{TRI-KC}, and the $O({\cal N})$ ferromagnetic coarsening in the infinite 
${\cal N}$ limit~\cite{Chcuyo} and including $1/{\cal N}$ corrections~\cite{Sollich-ON}
with the aim of checking predictions from the time-reparametrization 
invariance scenario of glassy dynamics~\cite{TRI-review}.

\begin{figure}
\vspace{0.3cm}
\includegraphics[width=0.43\textwidth]{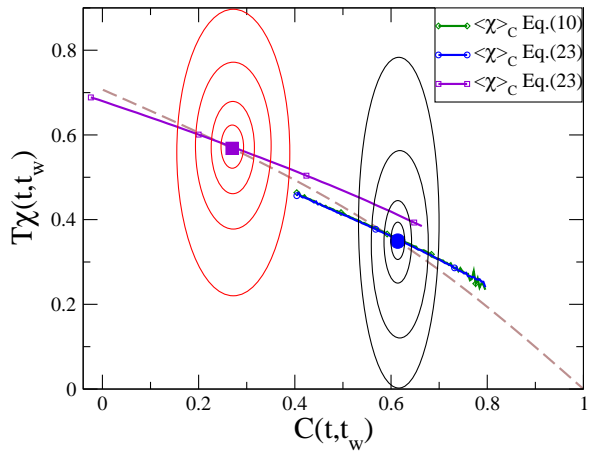}
\vspace{2cm}
\caption{
The procedure used to compute $\langle \chi \rangle _C
  (t,t_w)$ is illustrated for the $1d$ Ising model quenched to $T=0$.
  The dashed line is the curve $\widehat \chi (\langle C \rangle)$ and
  the bold symbols are the points of coordinates ($\langle C \rangle$,
  $\langle \chi \rangle$) for the two cases considered.  For a given
  choice of $t_w$ and $t$, that in the figure are $t_w=10$ and $t=20$
  (right) or $t=100$ (left), in units of Monte Carlo steps (MCs), one
  computes the joint probability distribution $P(C,\chi)$.  Some
  contour levels are sketched as ellipsoidal curves in the figure.  The
  average of $\chi$ at fixed $C$ (along the vertical direction) is
  then calculated. By varying $C$ one obtains the solid curves
  depicted in the figure. In the case $t=20$ (right) we show the
  restricted averaged integrated response obtained using 
  Eq.~(\ref{silly}) or by means of the field-free method based on
  Eq.~(\ref{alg}), as discussed in the text. The two curves superpose
  and are almost indistinguishable.}
\label{fig:scatter}
\end{figure}

The analysis of the statistics of two-time fluctuations
can be more conveniently carried over if some fluctuations are damped
by performing a partial averaging of $\chi $, in the following referred to as 
{\it restricted average}. Specifically, we introduce
the quantity 
\be
\langle \chi \rangle_C(t,t_w) =
\lim_{h\to 0} \frac{1}{N} \sum_i \frac{
  \langle (s^h_i(t)-s_i(t)) \epsilon_i \rangle _C}{h}
\label{silly}
\ee 
where the
{\it global} average $\langle \dots \rangle$ of Eq.~(\ref{eq:discrete-r})
is replaced by a {\it restricted} average $\langle \dots \rangle_C$ over configurations
with a given value of $C$ as sketched in
Fig.~\ref{fig:scatter}. As will be explained in the following,
considering restricted averages greatly simplifies the
analysis still retaining some basic physical information
on the relation between $\chi $ and $C$.
Since, even for $t$ and $t_w$ fixed, $C$ is a fluctuating quantity, 
the restricted averaging procedure allows us to explore the behavior of $\langle \chi \rangle_C$ 
as a function of $C$, similarly to what is done with the
globally averaged quantities $\langle \chi\rangle $ and $\langle
C\rangle$ by varying $t$ and $t_w$ (notice that, differently from globally averaged
quantities, the plot of $\langle \chi \rangle_C$ vs $C$ has also a
negative branch). The physical idea inspiring this
analysis is that the dependence of $\langle \chi \rangle_C$ on $C$ should bear the
same information as $\widehat \chi (\langle C\rangle)$ on some 
properties of the system, in particular on the
limiting value $T/X_\infty$, the effective temperature.


Let us state the problem more precisely.  Generally $\langle \chi \rangle_C$ depends
on $C$, $t$ and $t_w$ (see Fig.~\ref{fig:scatter}). Taking advantage of the monotonic
properties of $\langle C\rangle$ and $\langle \chi\rangle$ as functions of times, 
one can replace the temporal dependencies in
favor of $\langle C\rangle$ and $\langle \chi\rangle $. 
The restricted averaged quantity  $\langle \chi \rangle_C$ can then be recast as $\langle \chi\rangle_C
= \widehat \chi_C (C,\langle C\rangle,\langle \chi\rangle)$.    
In the limit $t_w\to \infty$, the dependence on $\langle \chi\rangle $ can
be dropped since in the models considered $\langle \chi\rangle$ approaches $\widehat \chi (\langle C\rangle)$, 
see Eq.~(\ref{eq:limit_chi}), and it becomes redundant. Hence 
\be 
\lim _{t_w\to \infty} \langle \chi \rangle_C(t,t_w)=
\lim _{t_w\to \infty} \langle \chi \rangle_C(C,\langle C\rangle)
\; . 
\label{once}
\ee
Now, it is convenient to extract a factor 
$\widehat \chi (\langle C\rangle)$ -- the integrated  fully averaged  linear response--
from the right-hand-side, and to use the {\it natural}
variable $C-\langle C\rangle$ describing the fluctuations of $C$ around the
average to write:
\be
\lim _{t_w\to \infty} \langle \chi \rangle_C(t,t_w)=
\widehat \chi (\langle C\rangle) \; \langle \chi \rangle_C(C-\langle C\rangle,\langle C\rangle)
\;   
\label{twice}
\ee
[for simplicity we use the same symbols for the two different functions 
$\langle \chi \rangle _C$ in Eqs.~(\ref{once}) and (\ref{twice})]. 
Equation~(\ref{twice}) is simply a rewriting of  $\langle \chi \rangle_C(t,t_w)$
in terms of the more {\it natural } variables $C$ and $\langle C\rangle$.
From this point on we proceed by using some physically motivated assumptions,
the soundness of which will be tested in Sec. \ref{restrict}. 
Let us notice first that in Eq.~(\ref{twice}) we are left with a two-parameter
dependence in $\langle \chi \rangle_C$, while $\widehat \chi (\langle C\rangle)$ depends on a single
parameter. The guiding idea of an analogy between $\langle \chi \rangle_C(t,t_w)$ and
$\widehat \chi(\langle C\rangle)$ discussed above
suggests that $C$ and $\langle C\rangle$ may enter 
$\langle \chi \rangle_C$ in a particular combination, thus reducing the number of
parameters to one, as for $\widehat \chi(\langle C\rangle)$.  Since $C$ has the 
upper bound $C(t,t)=1$, 
the natural scale of fluctuations is $1-\langle C\rangle$. Therefore, we make
the following scaling {\it Ansatz} 
\be 
\lim _{t_w\to \infty} \langle \chi \rangle_C(t,t_w)=
\widehat \chi (\langle C\rangle) \; f\left (\frac{C-\langle C\rangle}{1-\langle C\rangle}\right
),
\label{scaling}
\ee
where $f(x)$ is a scaling function from which, we conjecture, one can 
extract $X_\infty$ in the region $C=0$. 
In this paper we check this conjecture in 
the Ising model in $d=1,2,3$
quenched to the critical temperature or below. We find that the
scaling form (\ref{scaling}) is verified, and that {\it the slope of $f(x)$}
evaluated at $x_0=-\langle C\rangle/(1-\langle C\rangle)$, the 
$x$ value that corresponds to $C=0$, yields the limiting $X_\infty$:
\be
-\frac {df(x)}{dx}\mid _{x=x_0} = X_\infty 
\; . 
\label{main}
\ee
This claim is done asymptotically and we discuss its implications and how to 
take this limit in the body of the paper.

This Article is organized as follows. In Sec.~\ref{overview} we overview 
what is known about the scaling behavior of $\langle \chi \rangle $ and $\langle C\rangle$ 
in quenched ferromagnetic systems.
In Sec.~\ref{restrict}, after defining the restricted average and the methods used to
measure it, we check the validity of the {\it Ansatz}~(\ref{scaling}) and~(\ref{main}) 
in various systems. Specifically, in Sec.~\ref{d1} we consider the Ising
model in $d=1$ quenched to $T=0$ in an independent interface
approximation (Sec.~\ref{noninteract}) and by means of numerical simulations (Sec.~\ref{interact}). 
In Sec.~\ref{Tc} we then study the
Ising model in $d=2,3$ quenched to $T_c$ with an analytical Gaussian approximation (Sec.~\ref{GaussTc}) and numerically (Sec.~\ref{numTc}). 
Finally, the case of a quench of the $d>1$ Ising model below $T_c$ is
considered in Sec.~\ref{secbelow}. We discuss the results, the relations with
the reparametrization invariance symmetry and some open problems in
Sec.~\ref{disc}.

\section{The averaged linear response and correlation in non-disordered coarsening systems} \label{overview}

Before entering the field of fluctuations, it is useful to overview
the pattern of scaling behavior of the fully averaged 
quantities $\langle \chi\rangle $ and
$\langle C\rangle$ which are quite well understood in the relatively simple
case of clean (without quenched disorder) ferromagnetic
systems.

In the case of quenches to the critical point of a scalar ferromagnetic model
the averaged two-time functions satisfy the scaling forms
\begin{eqnarray}
\langle C\rangle(t,t_w) &=& t_w^{-a} \; g(t/t_w,t_0/t_w)
\label{scalccrit}
\; , 
\\
\langle \chi\rangle (t,t_w) &=& \chi ^{eq}-t_w^{-a}\; \overline g(t/t_w,t_0/t_w)
\; ,
\label{scalchicrit}
\end{eqnarray}
where $\chi ^{eq}=\beta $ is the static equilibrium
susceptibility, $a=0.115$ in $d=2$~\cite{a-d2,a-d3} and $a=0.506$ in $d=3$~\cite{a-d3}. 
In Eqs.~(\ref{scalccrit}) and (\ref{scalchicrit}) $t_0$ is a microscopic 
time needed to regularize $\langle C\rangle$ and $\langle \chi\rangle $ at 
$t/t_w=1$ ensuring $\langle C\rangle(t,t)=1$
and $\langle \chi\rangle (t,t)=0$. In the asymptotic limit
$t_w\to\infty$ for any $t/t_w$ fixed the averaged correlation 
vanishes and the integrated linear response reaches the equilibrium
value $\chi ^{eq}=\beta$ due to the $t_w^{-a}$ prefactors.
Equations~(\ref{scalccrit}) and (\ref{scalchicrit}) imply \cite{Godreche-Luck} 
\be
T\widehat  \chi (\langle C\rangle)=1-\langle C\rangle
\; ,
\label{fdtcrit}
\ee 
and for the explicit forms of $g$ and $\overline g$ found, 
interestingly, the limiting slope $X_\infty$ 
takes a non-trivial universal value
$X_\infty <1$, that if interpreted as yielding an 
inverse effective temperature, 
$T/X_\infty$, signals the presence of a higher effective temperature 
than the bath one in the peculiar limit of Eq.~(\ref{xinfty}). 
The value of $X_\infty$ has been determined using field theoretical techniques
up to second order in $\epsilon =4-d$~\cite{Caga}; one finds $X_\infty=0.429(6)$ and 
$X_\infty=0.30(5)$ in $d=3$ and $d=2$, respectively. 
Numerical calculations yield 
$X_\infty\simeq 0.4$ in $d=3$~\cite{Godreche-Luck} and 
$X_\infty\simeq 0.33-0.34$ in $d=2$~\cite{Xinfty-d2}.  

For quenches below the critical point, two time quantities split
up \cite{split} into a stationary (quasi-equilibrium) part and an
aging part
\begin{eqnarray}
\langle C\rangle(t,t_w) &=& \langle C\rangle^{st}(t-t_w)+\langle C\rangle^{ag}(t,t_w)
\; , 
\label{splitc}
\; \\
\langle \chi\rangle(t,t_w) &=& \langle \chi \rangle^{st}(t-t_w)+\langle \chi \rangle^{ag}(t,t_w)
\; ,
\label{splitchi}
\end{eqnarray}
with $\langle \chi \rangle^{st}$ and $\langle C \rangle^{st}$ related by the equilibrium FDT.
This means that $\langle \chi \rangle^{st}$ saturates to the static
susceptibility $\chi^{eq}=\beta (1-M^2)$, 
$M$ being the equilibrium magnetization density below $T_c$, 
in the finite characteristic
time of the equilibrium state. 
The aging parts obey the scaling forms~\cite{Corberi-review,aging-sub-critical}
\begin{eqnarray}
\langle C\rangle^{ag}(t,t_w)&=&g(t/t_w)
\; ,
\label{scalcbelow}
\; \\
\langle \chi \rangle^{ag}(t,t_w)&=&t_w^{-a}\overline g(t/t_w)
\; , 
\label{scalchibelow}
\end{eqnarray}
for large $t_w$. Above the lower critical dimension $d_L$ one has $a>0$,
implying $\lim _{t_w\to \infty}\langle \chi\rangle(t,t_w)=\langle \chi \rangle^{st}(t-t_w)$
and
\begin{eqnarray} 
T\widehat \chi (\langle C\rangle) =   \left \{ \begin{array}{ll}
        1-\langle C\rangle  \qquad \quad \, $for$ \qquad \langle C\rangle> M^2 \; ,  \\
        1-M^2   \qquad\;\;\;\;  $for$ \qquad \langle C\rangle\le M^2 \; . 
        \end{array}
        \right .
        \label{R1}
\end{eqnarray} 
Therefore $X_\infty =0$.
Let us stress that $\widehat \chi (\langle C\rangle)$ is determined
by the stationary part of the response function only, with
the aging parts producing finite time corrections
which depend on $\langle C\rangle$ and $t_w$. 

The same scaling structure (\ref{splitc})-(\ref{scalchibelow})
holds in the case of systems at the lower critical dimension quenched
to $T=0$, with the difference that the stationary parts vanish \cite{nota}
and $a=0$. This implies that $\widehat \chi (\langle C\rangle)$ is, in this case, 
a property of the
aging regime. Its exact computation yields a non-trivial form with
$X_{\infty}=1/2$~\cite{Xinfty-d1}. 

\section{Results for $\langle \chi \rangle_C$}\label{restrict}

In this section we study the behavior of the restricted averaged response
$\langle \chi \rangle_C$ in different systems. 

Reaching the vanishing applied field limit in the calculation of the
response function is a difficult task that has been discussed in
numerous studies of the fully averaged response. A variety of methods that
avoid applying a field and transform the globally averaged linear
response into correlation functions have been proposed and
tested~\cite{chatelain-ricci,algo}.  Elaborating on these ideas, we argue
that not only $\langle \chi \rangle$, but also the restricted average,
$\langle \chi \rangle_C$, can be computed over unperturbed
trajectories. Actually, following the same line of reasoning exposed
in~\cite{algo} but taking at the end of the calculation just the
restricted average over the $\chi$'s leads to 
\be 
2T\langle \chi
\rangle_C (t,t_w)=1-C(t,t_w)+\frac{1}{N}\sum _{i=1}^N \langle
s_i(t)\sum _{t'=t_w}^t B_i(t') \rangle _C
\ , 
\label{alg}
\ee
where 
\be
B_i=2s_i w_i 
\ , 
\ee
$w_i$ is the transition rate for flipping the spin $s_i$ and
$t'$ runs over elementary moves.
Notice that this relation between the response function and unperturbed
quantities is only valid for the averages (global or restricted) 
of $\chi $, while the relation between the right hand side of
Eq.~(\ref{alg}) and the fully fluctuating quantity $\chi $
is not, in principle, known. The availability of the fluctuation-dissipation 
relation (\ref{alg}), therefore, is one of the
great advantages of dealing with restricted averages instead of
considering directly $\chi $.

In numerical calculations one evolves many different realizations
of the system up to time $t$ grouping them according 
to the value of the fluctuating global overlap $C$.
$\langle \chi \rangle _C$ can then be computed in two
equivalent ways:

i) At time $t_w$ a replica of the system is created on which 
the perturbation is switched on, and the fluctuating quantity
$\chi $ is computed through Eq.~(\ref{eq:discrete-r}).  
With the set of points 
$(C,\chi)$ one then computes the restricted average $\langle \chi \rangle_C$ by 
averaging the $\chi$'s over all instances with the same $C$,
as described in Fig. \ref{fig:scatter}. A dependence on the magnitude of the 
applied field remains and one is interested in the $h\to 0$ limit.

ii) The right hand side of
Eq.~(\ref{alg}) (which is itself a fluctuating quantity) is computed
on the unperturbed trajectory and hence averaged over realizations with the same $C$.
There is no applied field in this case.

As shown in Fig. \ref{fig:scatter} the two methods give identical
results, but the second one is much more efficient computationally with,
moreover, the built-in limit $h\to 0$.  In the following, therefore,
we shall compute $\langle \chi \rangle _C$ using Eq.~(\ref{alg}).

In the definitions above the $C$ and $\chi$ are summed over all spins
in the sample and this poses a problem.  The use of large system
sizes ${\cal L}$, $N\gg 1$, is needed to avoid too important finite-size effects.
But for very large systems significant fluctuations of $C$ are rare
and one can only access tiny variations around $\langle C\rangle$.  In
order to avoid this difficulty we prefer to collect the statistics
over subsystems with, say, linear size $\ell$, each containing
$n=\ell^d$ spins, and then compute the restricted average through
Eq.~(\ref{alg}) with $n$ in place of $N$. Ideally, the coarse graining
length $\ell$ should be much larger than the lattice spacing $\ell \gg
\delta$.  Actually, the results of the analysis carried out in this
paper become independent of the coarse-graining length for large $\ell
$, but if $\ell $ is too small finite-size effects affect the
statistics of $C$ and $\langle \chi \rangle_C$ much in the same way as
a small ${\cal L}$ would introduce corrections with respect to the
thermodynamic limit ${\cal L}\to \infty$ in the global quantities.
Then, a convenient choice of the coarse graining length is to fix
$\ell $ to be sufficiently large so as to avoid significant finite size effects,
but not too large either, otherwise fluctuations would be
too rare. Since finite size effects in coarsening systems occur on
length-scales of the order of the growing length $L(t)$, a convenient and
realizable choice of $\ell$ is
\begin{equation}
\delta \ll L(t) \lesssim \ell \ll {\cal L}.
\end{equation}
A similar choice was made in \cite{Jaubert} and \cite{Aron} for the
study of fluctuations in the $3d$ Edwards-Anderson spin-glass and the
random field Ising model, respectively. The scaling of the correlation
fluctuations with the additional variable $\ell/L(t)$ during
coarsening was also discussed in \cite{Aron}.
In this paper we are not interested in checking this kind of scaling
but we simply use $\ell$ as a magnifying glass to tune the extent of 
fluctuations under investigation.

\subsection{$d=1$ Ising model quenched to $T=0$} \label{d1}

We  here focus on the one dimensional Ising model prepared in an initially
disordered configuration at infinite temperature and evolving with
Glauber dynamics at $T=0$.  The dynamics is then simply Brownian
diffusion of interfaces with annihilation upon meeting. The averaged
aging properties of this model are well
established~\cite{Godreche-Luck}. In two recent papers~\cite{Mayer} the
study of multi-point correlation functions revealed the existence of
dynamic heterogeneities. Here, we extend these studies to analyze the 
relation between the fluctuations of two-time quantities that once
averaged become the linear response and correlation. 

\subsubsection{Non interacting interfaces approximation}\label{noninteract}

Here we derive the expression for $\langle \chi \rangle_C$ in an
approximation in which interface annihilation is neglected but
interfaces travel a long distance in the interval $t-t_w$.  This is
expected to be correct in the regime $t_w\gg t_0$ (interfaces are very
distant in the sample) and $t/t_w >1$ (the chosen interface travels a
long distance in the $t-t_w$ interval and $C$ is significantly
different from $1$ although the limit $t/t_w\to \infty$ is not reached). 
Interestingly enough, 
we shall see that the results derived in this
limit capture the basic features of the fully interacting system for
all $t/t_w$, even for $t/t_w\gg 1$ when interactions between interfaces
should become important.

\begin{figure}
\vspace{0.3cm}
\includegraphics[width=0.43\textwidth]{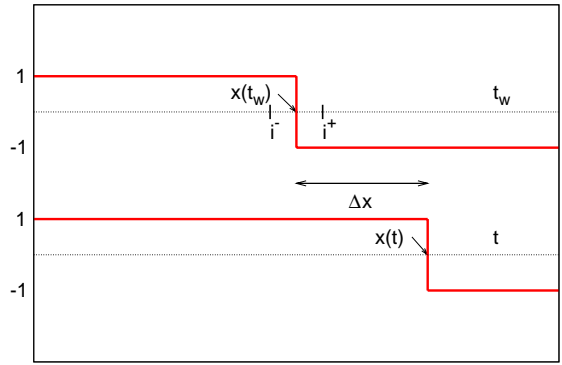}
\caption{Sketch of the interface motion in the $1d$ Ising model. 
The horizontal axis represents space and the vertical one the Ising spin configurations.
In the upper (lower) part of the figure the configuration at $t_w$ $(t)$ are shown.}
\label{fig:1dIM}
\end{figure}

For a single interface one can compute $\langle \chi \rangle_C$
exactly (see Appendix I for details). The result is 
\be 
2T\langle \chi \rangle_C
=1-C+\langle D \rangle _C,
\label{exact}
\ee 
where 
\be \langle D \rangle _C=\frac{2}{N}\sum_{n_c=0}^{t-t_w}\sum _{v_w=\pm 1} (2n_c+1+v_w) \; P_C(n_c,v_w)
\; . 
\ee
$n_c$ is the number of times the interface passes, in the interval
$[t_w,t]$, through its final position at time $t$, i.e. $x(t)$ (see Fig. \ref{fig:1dIM}). 
$v_w=+1$ ($v_w=-1$) if the first move of the interface (at time $t_w$) is
towards (away from) the final position, $x(t)$. In the sketch in Fig.~\ref{fig:1dIM}
this means that $v_w=1$ ($v_w=-1$) if the upper configuration moves to the right 
(left) in the next time-step. 
$P_C(n_c,v_w)$ is the
probability of finding a particular realization of $n_c$ and $v_w$ in the
restricted ensemble with a fixed value of $C$.  Notice that
$1-C=2\Delta x/N$ is entirely determined by the initial and final
position of the interface in terms of the distance $\Delta
x=x(t)-x(t_w)$ traveled.  A dependence on $x(t)$ and $x(t_w)$ also occurs
in $\langle D \rangle _C$ through $v_w$, because the probability of
moving toward the final position is always larger than that of moving
in the opposite direction.  However, for sufficiently large values of
$1-C$, when the interface has traveled a long distance, the sum is
completely dominated by the term $\sum _{n=0}^{t-t_w}
(2n_c+1)P_C(n_c)$, where $P_C(n_c)=\sum _{v_w=\pm 1}P_C(n_c,v_w)$.
Moreover, in the same limit, $P_C(n_c)$ itself does not depend on the
restriction $C$, because the position of the interface becomes
independent on the number of times the interfaces passes through $x(t)$.  In this limit,
therefore, $\langle D \rangle_C$ does not dependent on $x(t)$ and $x(t_w)$
and, hence, it is independent of $C$.  $\langle D \rangle_C$  can be determined by using the 
`sum rule' $\sum _C \langle \chi \rangle_C P(C)=\langle \chi\rangle$, where $P(C)$ is 
the probability distribution of $C$. 
This yields $\langle D\rangle _C=2T\langle \chi\rangle -1+\langle C\rangle$.  Hence one
has $T\langle \chi \rangle_C=T\langle \chi\rangle-(1/2)(C-\langle C\rangle)$.
Recalling~\cite{ontheconnection} that for a single interface
$T\langle \chi\rangle=1-C$, $\langle \chi \rangle_C$ can be cast in the scaling 
form~(\ref{scaling})
\be 
\lim _{t_w\to \infty} \langle \chi \rangle_C(t,t_w)
=
\widehat \chi (\langle C\rangle) \; 
f\left (\frac{C-\langle C\rangle}{1-\langle C\rangle}\right)
\qquad
\mbox{with}
\qquad
f(x)=1-\frac{1}{2} \ x 
\; . 
\ee 
The linear dependence of the function $f$ on its argument implies that its slope
is always equal to $-1/2=X_\infty$, which agrees with Eq.~(\ref{main}). 
This result, however, is expected to apply only 
within the limits of the approximation, that is to say, when $C$ is not too 
close to $1$ nor to $-1$. We  put this prediction to the numerical test in the next subsection. 

\subsubsection{Numerical results for the full model} \label{interact}

\begin{figure}
    \centering
   \rotatebox{0}{\resizebox{.5\textwidth}{!}{\includegraphics{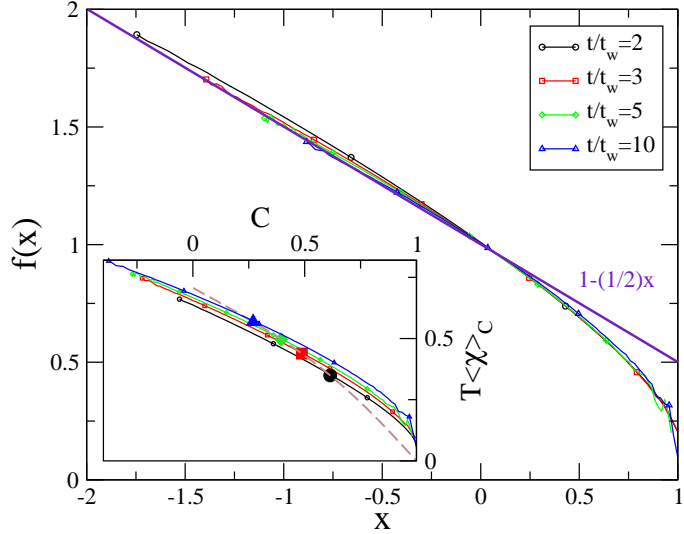}}}
   \vspace{2cm}
   \caption{(Color online.) Test of the scaling hypothesis (\ref{scaling}) in the
     Glauber dynamics of the one dimensional Ising model quenched to
     $T=0$ (using $N=10^6$ and $\ell =100,150,200,125$ respectively for the cases with
     $t/t_w=2,3,5,10$).  In the inset the quantity $\langle \chi \rangle_C$ is
     plotted against $C$ for different choices of $t/t_w$ given in the
     key and $t_w=10$.  The heavy symbols represent the points of
     coordinates ($\langle C\rangle,\langle \chi\rangle$) for the
     different choices of $t/t_w$, which lie on the fully averaged
     curve $\widehat \chi (\langle C\rangle)$ represented with a
     dashed line.  The curves $\langle \chi \rangle_C(C, \langle
     C\rangle)$ are close to each other but there is a small visible
     drift for increasing $t/t_w$ and they are not parallel. In the main panel $\langle \chi
     \rangle_C(C, \langle C\rangle)/\widehat\chi(\langle C\rangle)$,
     that defines the function $f(x)$ through Eq.~(\ref{scaling}), is
     plotted for the same set of data.  The solid (violet) line is the linear
     behavior, $f(x) = 1-x/2$, obtained in the non-interacting
     interface approximation of Sec.~\ref{noninteract} that very
     accurately describes the data in the negative region of
     $x$. See the text for a discussion.}
\label{scaling1d}
\end{figure}

In this section we study numerically the behavior of the fully
interacting model. Here, and in the following, we set the strength of
the interaction $J$ and the Boltzmann constant $k_B$ equal to 1
and time is measured in Monte Carlo steps (MCs).
After a few MCs the system enters the scaling regime in
which two-time averaged quantities depend only on the ratio $t/t_w$.  In the
inset of Fig.~\ref{scaling1d} we show the behavior of $\langle \chi
\rangle_C$ for different choices of $t/t_w$. The curves are clearly
distinct.  In the main part of the figure the collapse obtained by plotting
$\langle \chi \rangle_C/\langle \chi\rangle$ against $x=(C-\langle
C\rangle)/(1-\langle C\rangle)$ is demonstrated.  Moreover, the curve
with $t/t_w=10$ is numerically indistinguishable from $f(x)=1-x/2$
for all $x\le 0$. Upon decreasing $t/t_w$ one can notice a small
residual dependence on $t/t_w$, the larger the smaller $t/t_w$.
This is due to the fact that $n$ is finite. By increasing $n$ one can
check that the curves converge to a master-curve behaving as
$f(x)=1-x/2$ for $x\le 0$ (however averaging over larger boxes
reduces the extent of fluctuations and one can only study a small
region of $C$ values around $\langle C\rangle$).  This confirms the
validity of the scaling (\ref{scaling}).  

In conclusion, for $x\le 0$, $f(x)$ is
given by the non-interacting interface approximation.  Since
$x_0=-\langle C\rangle/(1-\langle C\rangle)<0$,
Eq.~(\ref{main}) holds and the limiting $X_\infty$ can be read from
the slope of $f(x)$ in the negative $x$ sector.
The fact that the approximation describes very accurately the data in the 
full $x \le 0$ sector is somehow surprising since neglecting interface interactions 
is hard to justify far from $x=0$.
The deviation of the function $f(x)$ from the linear shape in the 
region $x>0$ was to be expected since the approximation used to 
derive it, in particular the assumption $1-C$ significantly different 
from zero, is not respected. In particular, since for $C=1$ it must be
$\langle \chi \rangle _{C=1}\equiv 0$, $f(1)\equiv 0$ must hold and the
curve bends downwords with respect to the non-interacting interface
approximation.  

Let us stress that the non-trivial factor $\widehat \chi(\langle
C\rangle)$ in Eq.~(\ref{scaling}) has to be divided away to obtain
$X_\infty$: the slope of $\langle \chi \rangle_C$ against $C$ depends
on $\langle C\rangle$, it is not constant and differs from $X_\infty$ 
(see the inset in Fig. \ref{scaling1d}). Notice also that the
curve $\langle \chi \rangle _C(C,\langle C\rangle)$ is different from 
$\widehat \chi (\langle C\rangle)$ for any choice of $t,t_w$:
the two curves cross at $C=\langle C\rangle$ with a different slope.
We shall comment on the implications of these results on
time-reparametrization invariance in Sec.~\ref{disc}.

\subsection{Ising model in $d=2,3$ quenched to $T_c$} \label{Tc}

We consider now the Ising model quenched to the critical temperature
$T_c$ in three and two dimensions.  We first use a Gaussian approximation 
and later we present numerical simulations.

\subsubsection{Gaussian approximation} \label{GaussTc}

A Gaussian joint probability 
distribution of $C$ and $\chi$ reads
\begin{eqnarray}
P_{G}(C,T\chi) 
= 
\sqrt{\frac{\det A}{2\pi}} 
\exp\left\{ -\frac12 \; (\delta C, T \delta \chi) \ A \
\left( 
\begin{array}{c} 
\delta C\\
T \delta \chi
\end{array}
\right)
\right\}
\label{gaussa}
\end{eqnarray}
where $\delta C=C-\langle C\rangle$, $\delta \chi=\chi-\langle \chi\rangle$, 
$\langle C\rangle$ and $\langle\chi\rangle$ are the mean values that 
we introduce as time-dependent parameters, and the matrix $A$  
is   
\begin{eqnarray}
A &=& 
\left( 
\begin{array}{c}
V_C \;\;\; V_{C\chi} \\
V_{C\chi} \;\;\; V_\chi\\
\end{array}
\right)^{-1} 
=
\left( 
\begin{array}{c}
V_{\chi} \;\;\; -V_{C\chi} \\
-V_{C\chi} \;\;\; V_C\\
\end{array}
\right)
\;
\frac{1}{V_C V_{\chi} - V_{C\chi}^2}
\; . 
\end{eqnarray}
This PDF can only be an approximation in the critical dynamics of the 
finite dimensional Ising models. This forms is exact, instead, for 
quantities $C$ and $\chi$ that summed over all spins
in the $O({\cal N})$ model in the large ${\cal N}$
limit, as it has been analyzed in \cite{Chcuyo}.

Deviations from the Gaussian PDF are expected for finite ${\cal N}$ and 
finite $N$ systems. Annibale and Sollich recently studied the critical
dynamics of ferromagnetic spherical models with finite $N$. 
The detailed analysis of $1/\sqrt{N}$ corrections developed in  
\cite{Sollich-ON} showed that these can be treated perturbatively
at leading order for quenches at criticality and are thus amenable to 
analytic investigation.
We shall discuss the results in this paper and how they compare to 
ours in  the Discussion Section.

The restricted average yields 
\begin{eqnarray}
T \langle \chi\rangle_C^{G} &=&
\frac{T\int d\chi \ \chi \ P_{G}(C, T\chi)}{\int d\chi \ P_{G}(C,T\chi)}
=
T\langle \chi\rangle 
\left[ 1+ \frac{V_{C\chi}}{V_C} \, 
\frac{C-\langle C\rangle}{T\langle \chi\rangle} 
\right]
\; .  
\label{gauss1}
\end{eqnarray}
(Note that $\int dC \ \langle \chi\rangle_C \neq \langle \chi\rangle$.)
Recalling that for large $t_w$ the FDT holds 
$T\langle \chi\rangle=1-\langle C\rangle$  in a critical quench, 
Eq.~(\ref{gauss1})
has the scaling form~(\ref{scaling}) proposed if the ratio
$V_C/V_{C\chi}$ is a constant. Interestingly enough, in this framework
$X_\infty =-V_C/V_{C\chi}$ would be given by a ratio of covariances.

The joint PDF of $C$ and $\chi$ depends on the value of $\ell$, the 
coarse-graining length over which these quantities are computed. Clearly, 
for $\ell \gg L(t)$ fluctuations become rare and are
sharply distributed around the mean values with Gaussian statistics. 
In this limit the Gaussian 
approximation should be very precise. However, this limit is not that 
interesting for our purposes (and for the utility of this type of measurement 
as a method 
to estimate $X_\infty$) since the extent of fluctuations is 
heavily suppressed. In the more interesting case $\ell \stackrel{>}{\sim} L(t)$ 
the joint PDF cannot be Gaussian but, still, as we shall see numerically, 
the approximation is of relatively good quality and yields a
good estimate of $X_\infty$.

We have checked the quality of the Gaussian approximation and 
its implications on 
the validity of both Eq.~(\ref{scaling}) and Eq.~(\ref{main})
by computing numerically $V_C\equiv\langle \delta C^2\rangle$ 
and $V_{C\chi}=\langle \delta C \ T\delta \chi\rangle $ in the 
critical quench of the Ising model in $d=2$. The angular brackets indicate 
here an average over the numerical distribution function.

\begin{figure}
    \centering
   \rotatebox{0}{\resizebox{.5\textwidth}{!}{\includegraphics{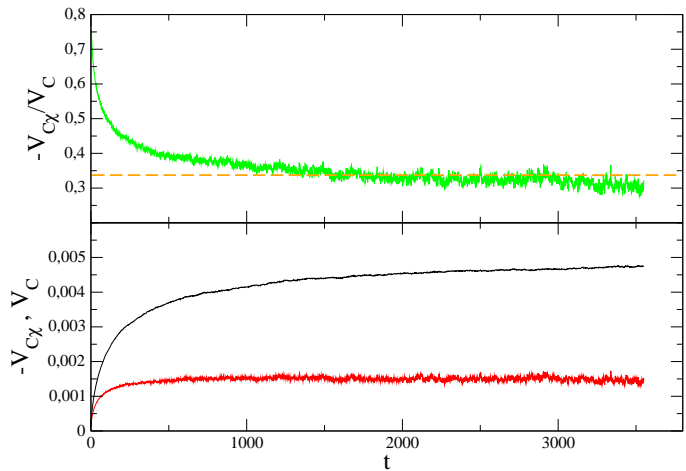}}}
   \vspace{2cm}
   \caption{(Color online.) In the lower panel the covariances $V_C$ (upper curve), 
and $-V_{C\chi}$ (lower curve)
are plotted against time for the Ising model in $d=2$ quenched to 
$T_c\simeq 2.269$. The system size is
${\cal L}=2\cdot 10^3$ and the coarse-graining length is
$\ell=85$.
In the upper panel the behavior of the ratio $-V_{C\chi}/V_C$ is shown. The dashed
line is the value of $X_\infty$.}
\label{varianze}
\end{figure}

In Fig.~\ref{varianze} we  display the time dependence of 
$V_C$ and $V_{C\chi}$. The curves initially depend on time, then 
reach (approximately) a plateau and their ratio, shown in the 
upper panel, a constant. At rather long times a time dependence
develops signaling that the Gaussian approximation
becomes less accurate, a fact that was to be expected since $L(t)$ 
increases and $\ell$ is no longer 
larger than $L(t)$. Focusing on the constant part of the ratio 
one notices that it  is very well close to the known value of $X_\infty$, indicated with a 
dashed horizontal line in the figure. 
These findings clearly substantiate our conjecture in 
Eqs.~(\ref{scaling}) and (\ref{main}).

Finally, let us add that the computation of $V_{\chi}$ at fixed $\ell$
(where $\chi $ is computed through Eq.(\ref {alg})) 
yields a monotonously increasing function 
diverging in the large-$t$ limit (not shown in Fig. \ref{varianze}). 
We shall comment on this issue in Sec.~\ref{disc}.

\subsubsection{Numerical results for the full model} \label{numTc}

\begin{figure}
    \centering
   \rotatebox{0}{\resizebox{.5\textwidth}{!}{\includegraphics{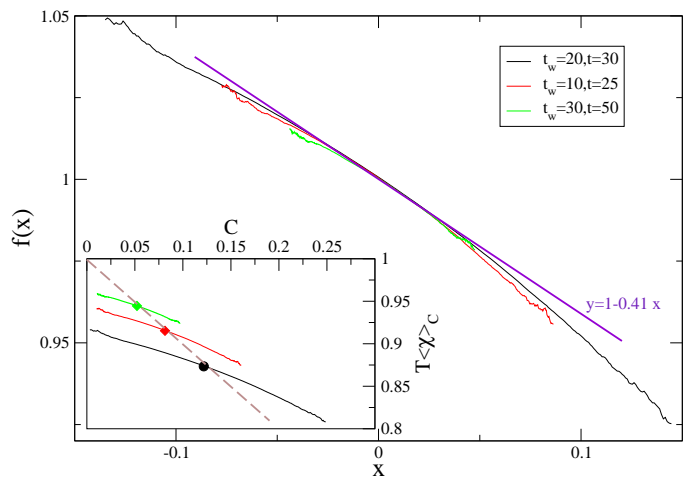}}}
   \vspace{2cm}
   \caption{(Color online.) Test of the scaling hypothesis
     (\ref{scaling}) in the critical dynamics of the three-dimensional
     Ising model ($T_c\simeq 4.511$). The system size is ${\cal
       L}=200$ and the coarse-graining length is $\ell=20,30,30$ for
     the cases ($t_w=20$, $t=30$), ($t_w=10$, $t=25$) and ($t_w=30$,
     $t=50$), respectively. In the inset the quantity $\langle \chi
     \rangle_C$ is plotted against $C$.  The heavy symbols represent the
     points of coordinates ($\langle C\rangle,\langle \chi\rangle$)
     for the different choices of $t,t_w$ given in the key and they
     all lie on the FDT curve $T\widehat \chi (\langle
     C\rangle)=1-\langle C\rangle$ (dashed line).  In the main part of the 
figure
     the function $f(x)$ defined through Eq.~(\ref{scaling}) is
     extracted from the same set of data.  The solid (violet) straight
     line is the expected small-$x$ behavior $f(x)=1-X_\infty x$ with
     $X_\infty \simeq 0.41$. Best fits of the slopes of $f$ in $x=0$
     yield $0.41$, $0.42$, $0.40$ for the cases ($t_w=20$, $t=30$),
     ($t_w=10$, $t=25$) and ($t_w=30$, $t=50$), respectively.}
\label{scalingcrit-d3}
\end{figure}

We start from the $d=3$ case quenched to the critical point.  In the inset in
Fig.~\ref{scalingcrit-d3} we show the plot of $\langle \chi \rangle_C$
versus $C$.  According to the scaling in Eqs.~(\ref{scalccrit}) 
and~(\ref{scalchicrit}) of the fully averaged quantities in the large
$t_w$ limit with fixed $t/t_w$ one has $\langle C\rangle\to 0$ and
$T\langle \chi\rangle \to 1$.  This implies that, in this limit, the
scaling~(\ref{scaling}) of the restricted averaged linear response is
lack of content since it becomes
$\langle \chi\rangle_C = f(C)$, just defining $f$.  However the truly
asymptotic limit cannot be reached numerically and, for the values of
$t_w$ used in the simulations, one still finds a difference of order
$0.05-0.15$ (depending on the curve) between $\langle C\rangle$ and
$0$, and $T\langle \chi\rangle $ and $1$. Despite these important
pre-asymptotic corrections in the independent behavior of $\langle
C\rangle$ and $T\langle \chi\rangle $ they still satisfy the
asymptotic relation~(\ref{fdtcrit}) with good precision.

In Fig.  ~\ref{scalingcrit-d3}
we test the scaling~(\ref{scaling}) for the pre-asymptotic
dynamics.  The collapse is excellent in
the region $x \simeq 0$ and looses quality for larger values of $|x|$,
probably due to too important finite $t_w$ corrections.  In order to
check the conjecture~(\ref{main}) we evaluate the slope of $f(x)$ in
$x=0$.  Since $x_0\to 0$ for $\langle C\rangle\to 0$ this procedure is
asymptotically equivalent to measuring the slope in $x_0$, with the
great advantage of using the region where the scaling~(\ref{scaling})
is well obeyed and where we have the best statistics.  The value of
the slope measured in this way yields similar values for all the
curves, all of which in remarkable agreement with $X_\infty \simeq
0.40 - 0.43$ as reported in the literature~\cite{Caga,Godreche-Luck}.
Note that in the limit $\langle C\rangle \to 0$ the prefactor
$T\widehat \chi (\langle C\rangle)$ tends to one and the slope of $f$ becomes
the slope of $T\langle \chi\rangle_C$ that coincides with the slope of
$T\widehat \chi (\langle C\rangle)$. The interesting regime has been
compressed to a single point in the $T\widehat \chi (\langle
C\rangle)$ graph but it opens up in the fluctuation analysis. The
local slopes of the fluctuating relation at fixed times and the fully
averaged relation using time as a parameter coincide.

\begin{figure}[h]
   \rotatebox{0}{\resizebox{.5\textwidth}{!}{\includegraphics{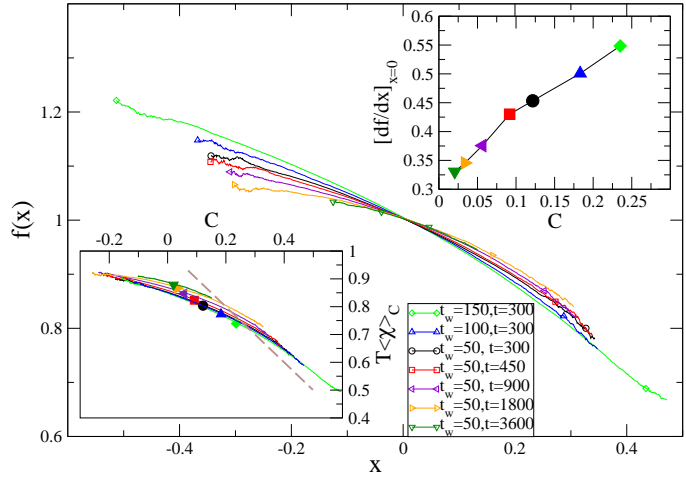}}}
   \vspace{2cm}
   \caption{(Color online.) Test of the scaling hypothesis
     (\ref{scaling}) in the critical dynamics of the two-dimensional
     Ising model ($T_c\simeq 2.269$). The system size is
     ${\cal L}=2\cdot 10^3$ and the coarse-graining length is
     $\ell=60$ except for the cases with 
     $(t_w,t)=(150,300)$ and $(t_w,t)=(50,3600)$ for which 
     $\ell =50$ and $\ell=85$ were respectively used.
     In the lower inset $T\langle \chi \rangle_C$ against
     $C$ is shown for different choices of $t$ and $t_w$ given in the key.  The
     heavy symbols represent the points of coordinates ($\langle
     C\rangle,\langle \chi\rangle$) and they fall away from the FDT
     curve $\widehat \chi (\langle C\rangle)=1-\langle C\rangle $ (dashed line).  In
     the main part of the figure the function $f(x)$ defined through
     Eq.~(\ref{scaling}). The upper inset shows the behavior of $df/dx \mid
     _{x=0}$ as $C=\langle C\rangle$ is varied. The trend seems
     to indicate $\lim_{C\to 0} df(x)/dx = -X_\infty\simeq -0.34$, as expected.  }
\label{scalingcrit-d2}
\end{figure}

The situation is qualitatively similar in $d=2$, although
pre-asymptotic effects are stronger.  This is due to the value of the
exponent $a$ which regulates the scaling behavior in Eqs.~(\ref{scalccrit})
and (\ref{scalchicrit}), that, being smaller in $d=2$ than in $d=3$,
delays considerably the asymptotic convergence~\cite{a-d2}.
Indeed, in order to reduce the deviations of $\langle C\rangle$ and
$T\langle \chi\rangle $ from the asymptotic values to values that are
comparable to the ones found in the $d=3$ case we had to use much
longer times.  Moreover, differently from the three-dimensional case,
$\langle C\rangle$ and $T\langle \chi\rangle $ do not obey
Eq.~(\ref{fdtcrit}) pre-asymptotically as shown in the lower inset to
Fig.~\ref{scalingcrit-d2} by the fact that the heavy symbols fall away from
the dashed straight line.  Due to such larger pre-asymptotic effects
there is a residual time dependence in the slope of $f(x)$ in the
origin, as shown in the main part of Fig.~\ref{scalingcrit-d2}, and
the collapse of the curves is not as good in $d=3$. However,
the quality of the collapse improves as the asymptotic region
is approached (namely as $\langle C\rangle $ decreases). Actually, the curves
corresponding to the two smaller values of $\langle C\rangle$ exhibit
a good collapse in the central region $C\simeq \langle C\rangle$.
Moreover, the
slope slowly approaches the known value
$X_\infty=0.33-0.34$~\cite{Xinfty-d2}, which is reached in the
longest run.  We stress that also in this
case the curve $\langle \chi \rangle _C(C,\langle C\rangle)$ is
different from $\widehat \chi (\langle C\rangle)$ but, in the asymptotic
limit in which $T\widehat \chi(\langle C\rangle) \to 1$ the slope of 
$\langle \chi \rangle _C(C,\langle C\rangle)$ at $C\to 0$ coincides with the 
slope of $\widehat \chi (\langle C\rangle)$ at $\langle C\rangle \to 0$
and they are both given by 
$-X_\infty$.
\vspace{1cm}

\subsection{Ising model in $d=2$ quenched below $T_c$} \label{secbelow}

We consider now the behavior of a ferromagnetic system quenched below
the critical temperature. We restrict the analysis to a two-dimensional
case because the task is computationally demanding and we do not expect
qualitative differences in higher dimensions.

We assume that a splitting analogous to Eq.~(\ref{splitchi}) holds
also for the restricted averages, namely 
\be 
\langle \chi
\rangle_C=\langle\chi\rangle_C^{st}+\langle \chi\rangle_C^{ag} \; .
\ee 
Since $\langle \chi \rangle_C^{st}$ is an equilibrium contribution
it depends only on the time difference $t-t_w$. Working with fixed
$t/t_w$ in the limit $t_w\to \infty$ amounts to probe the large $t$
limit of $\langle\chi\rangle_C^{st}$. Since this is a static quantity computed in
an equilibrium state (although with a restricted average) it cannot
depend on the history, and hence neither on $C$. Therefore, in the
limit considered one has $\langle\chi\rangle_C^{st}=\chi^{eq}=\beta (1-M^2)$.

Next, we want to show that $\langle\chi\rangle_C^{ag} $ can be
neglected with respect to $\langle\chi\rangle_C^{st} $, similarly to
what happens for the fully averaged quantities.  In appendix II we
present a scaling argument showing that in the large $t_w$ limit
$\langle\chi\rangle_C =0$ for a quench to $T=0$. Since in this case
$\langle \chi \rangle_C =\langle\chi\rangle_C^{ag}$ and the effect of
a finite temperature is not expected to change significantly the
behavior of the aging contributions, this argument suggests that
$\langle\chi\rangle_C^{ag} $ can indeed be neglected for any $T<T_c$.
 
\begin{figure}[h]
    \centering
   \rotatebox{0}{\resizebox{.5\textwidth}{!}{\includegraphics{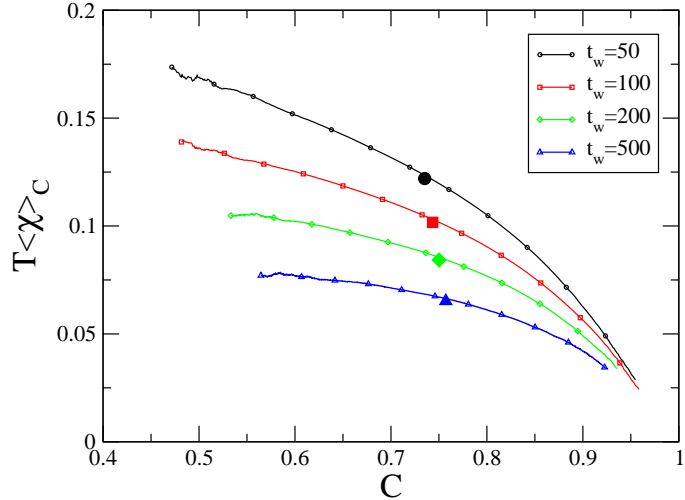}}}
   \caption{(Color online.) The quantity $\langle \chi \rangle_C$ is
     plotted against $C$ for the two-dimensional Ising model quenched
     to $T=1<T_c$ using $t/t_w=2$ and different choices of $t_w$. The
     system size is ${\cal L}=2\cdot 10^3$ and the coarse graining
     length is $\ell =50, 70, 100, 150$ for the cases with
     $t_w=50,100,200,500$, respectively.  The heavy symbols represent
     the point of coordinates ($\langle C\rangle,T\langle \chi\rangle$)
     for the different choices of $t_w$. The curves are
     slowly turning flat as $t_w$ increases implying that $X_\infty\to
     0$ as expected.}
\label{below}
\end{figure}

In the following we test this statement
numerically.  The most obvious way of computing
$\langle \chi\rangle_C^{ag}$ is by subtracting
$\langle\chi\rangle_C^{st}$ from $\langle \chi \rangle_C$. However
there is a by far more efficient way to compute $\langle\chi\rangle_C^{ag}$
that consists in
considering a modified dynamics in which flips in the bulk of domains
are prevented. Since the stationary contribution is given by the
reversal of spins well inside the domains this no-bulk-flip dynamics
isolates the aging behavior with the numerical advantage of
evolving only the small fraction of interface spins. This technique
has been thoroughly tested and used in studies of
$\langle \chi\rangle $ and $\langle C\rangle$~\cite{ontheconnection,NBF}. We have checked that
also for the restricted average response the no-bulk-flip kinetics yields the same results
as subtracting $\langle\chi\rangle_C^{st}$ from $\langle \chi \rangle_C$. The
results obtained with this kind of dynamics are shown in
Fig.~\ref{below}.  One concludes that, working with a fixed $t/t_w$
(for instance, the set of data obtained with $t/t_w=2$ are shown in the figure), for any
given value of $C$, $\langle \chi \rangle_C $ and its slope go to
zero.  This guarantees that for very long times $\langle\chi\rangle_C^{ag}$ can be
neglected with respect to $\langle\chi\rangle_C^{st}$ and hence one has 
$T\langle \chi \rangle _C (t,t_w) =1-M^2$
and $f(x)\equiv 1$, leading to a  
vanishing slope and $X_\infty=0$.  
Notice that the mechanism producing $T_{eff}=\infty$ in the 
full aging regime (and in consequence $X_\infty=0$) is
the same as for the global quantities, namely the fact that the
aging contribution vanishes asymptotically.

\vspace{2cm}

\section{Discussion} \label{disc}

In a series of papers it was claimed that time-reparametrization
invariance is the symmetry that controls dynamic
fluctuations~\cite{TRI-SG,TRI-KC, TRI-review} in aging systems with a
finite effective temperature, namely low temperature glassy cases.  
A consequence of
this proposal is that the fluctuating linear responses and correlation
functions computed over the same realization (in the same subsystem
of size $\ell$ with the same stochastic noise) should, 
in the $t_w\to\infty$ limit, the aging regime and the 
scaling limit $\delta \ll \ell \ll
{\cal L}$, be linked and satisfy
\begin{equation}
\chi^{ag}(t,t_w) = \widehat \chi^{ag}(C)
\qquad \mbox{with} \qquad C=C(t,t_w)
\label{eq:TRI}
\; . 
\end{equation}
Importantly enough, the function $\widehat\chi^{ag}(x)$ is the scaling
function of the global and fully averaged linear response in the aging
regime in which the latter is a non-trivial function $\widehat \chi^{ag} (\langle C\rangle)$ 
of the global and fully averaged correlation, see Eq.~(\ref{eq:limit_chi}).  
The requirement of having a finite effective temperature
translates into the fact that $ \widehat \chi^{ag}(C)$ does not vanish
asymptotically. The statement (\ref{eq:TRI}) is equivalent to saying
that the fluctuations along the curve $\widehat \chi^{ag}(C)$ are
massless while the ones that do not follow this direction are massive
and can hence be eliminated by the coarse-graining in the scaling limit. 
Time-reparamentrization invariance is a symmetry that is expected to 
develop asymptotically. At finite times $\ell$ should be scaled with a 
growing correlation length that diverges asymptotically.
Numerical tests of
this proposal in the low temperature dynamics of the $3d$
Edwards-Anderson model~\cite{TRI-SG}, some kinetically constrained
systems~\cite{TRI-KC}, Lennard-Jones mixtures~\cite{Castillo} 
and disordered elastic lines~\cite{Joseluis}
yielded encouraging results.

The search for time-reparametrization symmetry in the $O({\cal N})$ coarsening
model in the large ${\cal N}$ limit showed that this symmetry is not fully
realized in this case; it is, instead, reduced to global rescalings of
time, $t \to \lambda t$ with $\lambda$ a constant parameter~\cite{Chcuyo}.  This result suggested that
dynamic fluctuations in coarsening problems might follow a different
rule although the question remained as to whether the reduction of
time-reparametrization invariance to time-rescaling was a pathology of
the large ${\cal N}$ limit.

\begin{figure}[h]
    \centering
   \rotatebox{0}{\resizebox{.5\textwidth}{!}{\includegraphics{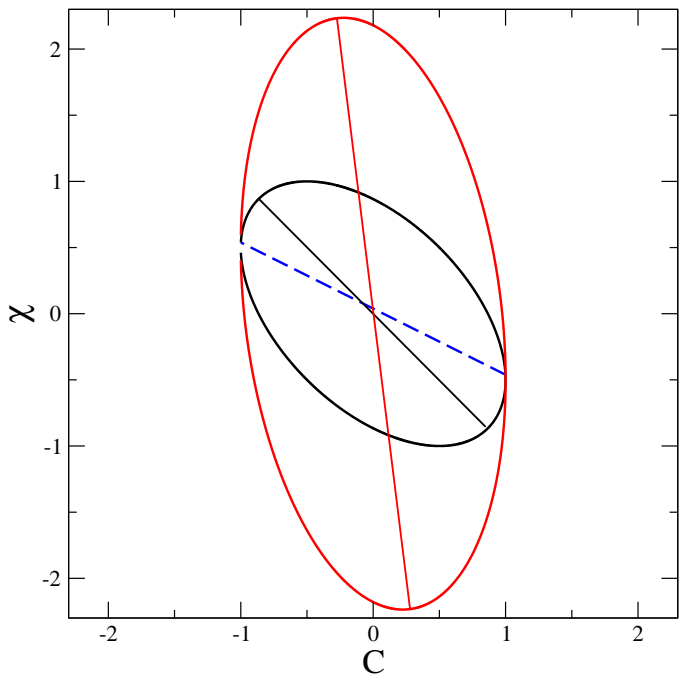}}}
   \caption{(Color online.) The ellipse defining the surface of constant probability
   $P(C,T\chi)=const.$ in the Gaussian approximation (\ref{gaussa}).
   The two ellipses have $V_C=2 V_{C\chi}=1$ while $V_{C\chi}=1$ and $V_{C\chi}=5$
   for the inner and the outer ellipse respectively. The slope $\tan \theta=-V_{C\chi}/V_C$
   (dashed line) is the same for the two ellipses, while the slope $\tan \phi $ of the major axis 
   (continuous straight lines) is different in the two cases.}     
\label{ellipse}
\end{figure}

Annibale and Sollich recently started the study of critical 
out of equilibrium dynamic fluctuations
by analysing the ferromagnetic spherical model with finite number of spins, $N$, 
including $1/\sqrt{N}$ 
corrections \cite{Sollich-ON}. The joint PDF of the global (summed over 
all spins in the system) $C$ and $T\chi$ 
is Gaussian for $N \to \infty$ and the contour levels 
are ellipses. For finite $N$ the fluctuations deviate from Gaussian statistics;
however,  at leading order in $1/\sqrt{N}$ the critical fluctuations 
can be treated perturbatively, and one recovers Gaussian statistics for 
$(C,T\chi)$ with elliptic contour 
levels that can be computed analytically. 
The principal axis of any of these ellipses forms an angle $\phi$ with the 
$C$ axis that is given by 
\begin{equation}
[\tan \phi ]^{-1}= \tan [(1/2) \ \mbox{atan} (2V_{C\chi}/(V_C-V_\chi))]
\ ,
\label{eq:phi}
\end{equation}
see Fig. \ref{ellipse}.
Notice that the angle $\phi $ depends
not only depends on  $V_C$ and $V_{C\chi}$ but also on
$V_\chi $.
 As discussed by 
Annibale and Sollich, if the time-reparametrization invariance 
scenario applied to critical dynamics, the angle $\phi$, that is a natural measure 
of the slope of the cloud, should yield
the fluctuation-dissipation ratio $X(\langle C\rangle)$
that relates the variations with time of the
{\it average} susceptibility and correlation, see Eq.~(\ref{eq:limit_chi}).
In particular, it should yield 
$X_\infty$ when the 
long $t$ limit is taken before the long $t_w$ limit.
The analytic computation of $V_C$, $V_\chi$ and $V_{C\chi}$ in $d<4$ 
showed that the angle is not related to $ X$ in any simple way. For  
small time differences the variances and co-variance are stationary 
but one does not recover the FDT slope $1$ from this calculation and 
in the opposite $t\gg t_w$ or $C\to0$ limit the angle tends 
to $\pi/2$ meaning that the ellipses stretch in the susceptibility 
direction. These results invalidate one consequence of 
time-reparametrization invariance and indicate that this symmetry 
does not develop asymptotically in the critical dynamics of the 
ferromagnetic spherical model at leading order in $1/N$.

In this paper we analyzed the linear-response/correlation fluctuations
in critical dynamics and coarsening in finite dimensional systems with
finite dimensional order parameter. 
The main point of this paper is to propose the use of  
restricted averages, in which the integrated 
linear responses are averaged over trajectories that have the same 
value of the fluctuating two-time function $C$, to study fluctuations 
in critical and sub-critical out of equilibrium dynamics.  We sudied 
the relation between restricted averaged susceptibility and 
fluctuation two-time function and we conjectured that it yields 
the asymptotic effective temperature $T/X_\infty$ relevant to 
critical dynamics. 
Although time-reparametrization invariance 
was not checked explicitly, our results bear some indication on the existence 
or not of such a symmetry in these cases.  
We summarize the results for different cases below.

We discuss  quenches to $T_c >0$ first. In these case, 
a Gaussian approximation is rather accurate if one uses coarse-graining 
lengths, $\ell$, that are significantly larger than the growing length $L(t)$.
Clearly, as time increases $L(t)$ goes beyond $\ell$ and the approximation 
has to be revised. 
The restricted average using the Gaussian PDF then naturally
provides a slope $\tan \theta =-V_{C\chi}/V_C$ that is different 
from $\tan\phi$, see Eq.~(\ref{eq:phi}). As shown
in Fig. \ref{ellipse}, this is the slope
of the line connecting the two points of the ellipse where $C$
takes the largest (smallest) value. This quantity does not depend on 
$V_\chi $ and was shown to be the one yielding $X_\infty$. Indeed, 
$V_{C\chi}$ and $V_C$ converge to a constant for long (but not too long)
$t$ and their ratio equals $-X_\infty$. Since $\theta $ yields $X_\infty $ 
and $\phi \neq \theta$ our results suggest that time reparametrization
invariance does not hold in this case. 
Note that despite the strictly 
Gaussian character of fluctuations in the large-$N$ spherical ferromagnetic 
model, the $V_C$, $V_\chi$ and $V_{C\chi}$ behave in a radically
different way from our determination. In particular, the ratio
between $V_{C\chi}$ and $V_C$ does not approach a constant in $d<4$.  

$V_\chi$ grows very fast as a function of time when the fluctuations are 
computed at fixed $\ell$. Therefore, the fluctuations of $\chi $ 
are very important and cannot be reduced -- as compared to the ones of
the corresponding correlation -- by using a convenient choice of
$\ell$. As found by Annibale and Sollich for 
the spherical  model, the axis of the cloud tends to turn parallel to the 
$\chi $ axis. This behavior is also at odds with what one would expect
if time-reparametrization invariance were obeyed, since the cloud should 
asymptotically lay parallel to the function $\widehat\chi^{ag}(x)$,
namely with a finite slope $X_\infty$.

\vspace{2cm}

The situation is different for the quench to $T=0$ in $d=1$.
The numerical data and
the independent interface approximation of the
kinetics of the Ising chain showed that $\langle \chi\rangle_C$ is not
equal to $\widehat \chi(C)$. Instead, we found that $\langle
\chi\rangle_C$ is given by a scaling function that is proportional to
the non-trivial function $\widehat \chi(C)$ multiplied by another
non-trivial function of $C$ and $\langle C\rangle$. These results
suggest that time-reparametrization symmetry may hardly be realized in this case and that the relation between the slope of the cloud and
$X_\infty$ is even more hidden than in the case of quenches to $T_c>0$. 
Nevertheless, despite the difference between $\langle
\chi \rangle _C$ and $\widehat \chi (\langle C\rangle)$, 
interestingly their
relation still encodes the limiting effective temperature 
$X_\infty$ through the scaling function $f$ defined from
$\langle\chi \rangle _C/\widehat \chi (\langle C\rangle)$.

For proper sub-critical coarsening the aging contribution to the
restricted averaged integrated linear response vanishes asymptotically
and, as for the fully averaged quantity, $T \langle \chi\rangle_C$
approaches the constant $1-M^2$. In a plot $T \langle \chi\rangle_C$
against $C$ one then just sees horizontal fluctuations but this is a
trivial consequence of the fact that $T \langle \chi\rangle_C$
approaches a constant. This result extends the one found in
\cite{Chcuyo} for the $O({\cal N})$ model in the infinite ${\cal N}$
limit to domain growth with finite dimensional order parameter.

The study of the same fluctuations could also be addressed
experimentally.  A recent study of the out of equilibrium relaxation
after a quench to the Fr\'eedericksz second-order phase transition
in a liquid crystal demonstrated that the fully averaged  correlation 
and linear response age and are linked by an FDR with an effective temperature
that is higher than the environmental one~\cite{Ciliberto}. 
The analysis of the fluctuations of these quantities and the restricted 
average proposed in this paper should shed light on the generality of 
our scaling hypothesis and the fact that $X_\infty$ could be accessed by 
studying fluctuations.

\vspace{1cm}
{\large {\bf APPENDIX I}}
\vspace{1cm}

Let us consider an interface $I$ located in $x(t')$ ($t_w\le t'\le t$),
namely $s_i=+1$ for $i<x(t')$ and $s_i=-1$ otherwise, see Fig.~\ref{fig:1dIM}. 
We denote with $i^-(t')$ and $i^+(t')$ the sites surrounding
$I$ (on the left and right respectively).  
One has 
\be
1-C=\frac{2\Delta x}{N},
\label{cconf}
\ee
where $\Delta x=| x(t)-x(t_w)| $. 
Let us now evaluate the term 
\be
\frac{1}{2}D(t,t_w)=\frac{1}{N}\sum _{i=1}^N s_i(t)\sum _{t'=t_w}^t \frac{1}{2}B_i(t')
\label{di}
\ee
appearing in Eq.~(\ref{alg}).
Let us stipulate that the velocity $v(t')$ is $+1$ ($-1$)
if the interface moves to the right (left). Notice that we are implicitly
assuming Metropolis-like transition rates (namely the interface
always moves). 
It is useful to introduce the {\it map of accelerations} $a_i$,
defined as follows.
Starting from $a_i\equiv 0$ an update occurs only
in the following cases:
\begin{itemize}
\item{When $I$ starts moving at $t'=t_w$: One changes
$a_{i^-(t_w)}\to a_{i^-(t_w)}+1$ ($a_{i^+(t_w)}\to a_{i^+(t_w)}-1$) if the $v_w=v(t_w)$
is positive (negative). Here and in the following all the $a_i$ 
not specifically mentioned remain unchanged.}
\item{When $I$ stops moving at $t'=t$: One updates
$a_{i^+(t)}=a_{i^+(t)}-1$ ($a_{i^-(t_w)}=a_{i^-(t_w)}+1$) if $v(t)$
is positive (negative).}
\item{When the velocity changes from $+1$ to $-1$ 
($-1$ to $+1$) in $x(t')$: One changes
$a_{i^-(t')}\to a_{i^-(t')}-1$, $a_{i^+(t')}\to a_{i^+(t')}-1$ 
($a_{i^-(t')}\to a_{i^-(t')}+1$, $a_{i^+(t')}\to a_{i^+(t')}+1$).}
\end{itemize}

At $T=0$ one has transition rates
$w_i(t')=1$ on sites $i^-(t'),i^+(t')$ and $w_i(t')=0$
elsewhere.
Then one has 
\begin{eqnarray} 
\frac{1}{2}B_i(t')=   \left \{ \begin{array}{rl}
        1    \qquad & $for$ \qquad i=i^- (t') \; , \\
        -1   \qquad & $for$ \qquad i=i^+ (t') \; , \\
        0    \qquad & $for$ \qquad i\neq i^-(t'),i^+(t').
        \end{array}
        \right .
        \label{bi}
\end{eqnarray} 
When $I$ moves these $\pm 1$ contributions
are seeded in the region traveled which will be
then summed up in $D(t,t_w)$.
If there are no accelerations (in the sense defined above)
all these contribution
sum up to zero in computing the integral over time 
in Eq.~(\ref{di}). Since contributions to the integral come only
from accelerations it is easy to prove that
one has 
\be
\frac{1}{2}D(t,t_w)=\frac{1}{N}\sum _i  s_i(t)\sum _{t'=t_w}^t a_i(t').
\label{doublesum}
\ee 
Most of the contributions to the sum over times cancel each other.
Let us indicate with $t_k$ the time when $I$ crosses $x(t)$ for the
$k$-th time. By definition $I$ changes direction an odd number of
times in every interval $[t_k,t_{k+1}]$.  Since the contributions due
to $I$ changing direction an even number of times without crossing
$x(t)$ cancel out in the sum over times of Eq.~(\ref{doublesum}), one
is left only with the accelerations at (say) the last velocity
reversal.  Then, for each interval
$[t_k,t_{k+1}]$ there is a contribution $+2$ ($-2$) if the interface
is on the right (left) of $x(t)$ which, once multiplied by $s(t)$
in Eq.~(\ref{doublesum}) gives a contribution $+2$.  In
conclusion, for $I$ crossing $x(t)$ $n_c$ times one has a contribution
$2n_c$ to $(1/2)D(t,t_w)$ plus the accelerations at $t_w$ and 
$t$.  Since the latter yield a contribution $+1$ one has
$(1/2)D(t,t_w)=[2n_c+1+v_w]/N$ and hence 
\be 
\langle D(t,t_w)\rangle_C=\frac{2}{N}
\sum _{n_c=0}^{t-t_w}\sum _{v_w=\pm 1}[2n_c+1+v_w]P_C(n_c,v_w), 
\ee
where $P_C(n_c,v_w)$ is the probability of finding a particular choice
of $n_c$ and $v_w$ in the restricted ensemble.

\vspace{1cm}
{\large {\bf APPENDIX II}}
\vspace{1cm}

\begin{figure}
    \centering
   \rotatebox{0}{\resizebox{.5\textwidth}{!}{\includegraphics{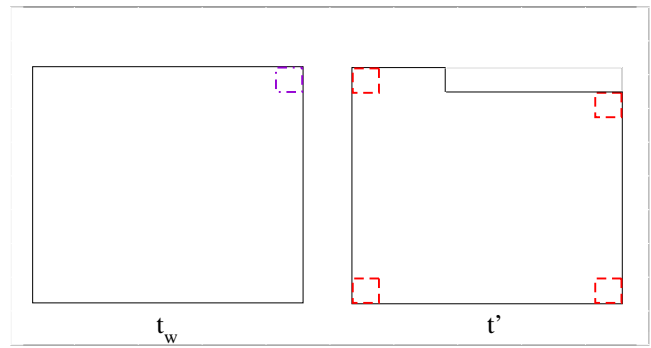}}}
    \caption{The relaxation of a domain with faceted interfaces at
     $T=0$. Left: The initial configuration, where only corner spins
     can be flipped. Right: Configuration at a generic time $t'$ with
     a kink. The four corner spins producing the relevant contribution
     to $D$ are evidenced.}
\label{scatola}
\end{figure}

In this section we develop a scaling argument to compute 
$\langle \chi \rangle_C$ in a quench to $T=0$ in $d=2$.
We shall assume Metropolis transition rates, as for $d=1$. 
We follow the scaling approach 
developed in~\cite{temp0}, which amounts to consider
the relaxation of a domain of (say) down spins which at time $t_w$ has a faceted interface,
as depicted in Fig.~\ref{scatola}. The process ends at
the final time $t_w+\tau $ when the domain has disappeared,
namely all its spins have been reversed.
The autocorrelation function of this process is
easily evaluated as $C(t_w+\tau,t_w)=1-2N_D/N$,
where $N_D$ is the initial number of spins in the
domain.
Let us evaluate the
quantity $(1/2)D(t_w+\tau,t_w)$ in Eq.~(\ref{di}), recalling that
$s_i(t)\equiv 1$.
At time $t'=t_w$ the only possible
moves are the flip of a corner spin. Let us assume for simplicity
that this is the top right, as shown in the left panel in 
Fig.~\ref{scatola}. This move generates a kink which performs a
random walk on the edge of the domain until it disappears
when it reaches the boundary on the left side (or another
anti-kink generated by the flipping of the spin on the
upper left corner). In this way the first row is eliminated,
and the process is then repeated until the domain 
disappears at time $t'=t_w+\tau$. At $T=0$, only spins
the flip of which do not increase the energy can be updated. Hence
at a generic time $t'$, the only 
contributions to $(1/2)D$ are those provided by the spins in
the corners or those surrounding kinks. The contribution
of the corners is always $-1$ while spins surrounding a kink
contribute $\pm 1$ ($-1$ for the spin belonging to the domain,
$+1$ for the other). As the dynamics proceeds, many of these contributions
are generated, that must then be summed up in $(1/2)D$.
In so doing, however, it is easy to realize that all the
contributions coming from the kinks sum up to zero, since
they always occur in pairs. Hence one is left with the
contributions from the corners only. As shown in the right panel
in Fig.~\ref{scatola}, topological reasons fix the number of
corners to be 4 at all times. The contribution $-1$ of these spins
lasts for all the time $\tau $ of the process.
Then one has $(1/2)D=-4\tau /N$.
The next step is to evaluate $\langle D \rangle _C$.
Since $C$ is univocally determined by $N_D$, 
computing $\langle D \rangle _C$ simply amounts to
determine the average time needed for a faceted
domain of $N_D$ spins to disappear with zero
temperature dynamics, namely 
$\langle D \rangle _C=-8\langle \tau \rangle_{N_D}/N$. 
In \cite{temp0} it was shown that 
$\langle \tau \rangle _{N_D}=N_D/4$.
Then one has $\langle D \rangle _C=-2N_D/N=-(1-C)$,
and from Eq.~(\ref{alg}):
\be
\langle \chi \rangle_C(t,t_w)=0
\; . 
\ee

\vspace{1cm}

We thank C. Aron, S. Bustingorry, C. Chamon, J. L. Iguain, 
M. Zannetti for useful discussions.
F. C. acknowledges financial support from PRIN 2007JHLPEZ 
({\it Statistical Physics of Strongly correlated systems in Equilibrium
and out of Equilibrium: Exact Results and Field Theory methods}) and from
CNRS and thanks the 
LPTHE Jussieu for hospitality during the preparation of this work. 
L. F. C. is a member of Institut Universitaire de France.


\vspace{1cm}

\begin{thebibliography}{99}

\bibitem{Cuku} L. F. Cugliandolo and J. Kurchan, Phys. Rev. Lett. {\bf 71}, 173 (1993);
J. Phys. A {\bf 27},  5749 (1994). 

\bibitem{Corberi-review}
F. Corberi, E. Lippiello, and M. Zannetti, J. Stat. Mech. (2007) P07002.

\bibitem{Caga} P. Calabrese and A. Gambassi, J. Phys. A {\bf 38}, R133 (2005). 

\bibitem{Godreche-review}
C. Godr\`eche and J.-M. Luck,
J. Phys. Cond. Matter {\bf 14}, 1589 (2002).

\bibitem{Crri} A. Crisanti and F. Ritort,
J. Phys. A  {\bf 36}, R181 (2003). 

\bibitem{Cukupe} L. F. Cugliandolo, J. Kurchan, and L. Peliti, 
Phys. Rev. E {\bf 55}, 3898 (1997). 

\bibitem{fmpp}
S. Franz, M. M\'ezard, G. Parisi, and L. Peliti,
Phys. Rev. Lett. {\bf 81}, 1758 (1998);
J. Stat. Phys. {\bf 97}, 459 (1999).

\bibitem{Sollich-fields} 
A. Annibale and P. Sollich, J. Phys. A {\bf 39}, 2853 (2006). 

\bibitem{Godreche-Luck} 
C. Godr\`eche and J.-M. Luck, 
J. Phys. A {\bf 33}, 9141 (2000).

\bibitem{Sollich-Xinfty} P. Mayer, L. Berthier, J. P. Garrahan, and P. Sollich, 
Phys. Rev. E {\bf 68}, 016116 (2003). P. Sollich, S. Fielding, and P. Mayer, J. Phys. Cond. Matt.
{\bf 14}, 1683 (2002). A. Garriga, P. Sollich, I. Pagonabarraga, and F. Ritort, 
Phys. Rev. E {\bf 72}, 056114 (2005). 

\bibitem{Caga-Xinfty} 
P. Calabrese and A. Gambassi, J. Stat. Mech. P07013 (2004). 

\bibitem{Corberi-Xinfty}
F. Corberi, E. Lippiello, and M. Zannetti, 
Phys. Rev. E {\bf 68}, 046131 (2003).  
 
\bibitem{TRI-SG} H. E. Castillo, C. Chamon, L. F. Cugliandolo, and M. P. Kennett, 
Phys. Rev. Lett. {\bf 88}, 237201 (2002). 
C. Chamon, M. P. Kennet, H. E. Castillo, and L. F. Cugliandolo,
Phys. Rev. Lett. {\bf 89}, 217201 (2002).  
H. E. Castillo, C. Chamon, L. F. Cugliandolo, J. L. Iguain, and M. P. Kennett, 
Phys. Rev. B {\bf 68}, 134442 (2003). 

\bibitem{Jaubert} L. D. C. Jaubert, C. Chamon, L. F. Cugliandolo and M. Picco,   
J. Stat. Mech. (2007) P05001.

\bibitem{TRI-KC} C. Chamon, P. Charbonneau, L. F. Cugliandolo, D. R. Reichman, and M.
Sellitto, J. Chem. Phys. {\bf 121}, 10120 (2004). 

\bibitem{Chcuyo} C. Chamon, L. F. Cugliandolo and H. Yoshino,
J. Stat. Mech. P01006 (2006). 

\bibitem{Sollich-ON} A. Annibale and P. Sollich, arXiv:0811.3168.

\bibitem{TRI-review} C. Chamon and L. F. Cugliandolo, 
J. Stat. Mech. P07022 (2007).

\bibitem{a-d2} 
F. Corberi, A. Gambassi, E. Lippiello and M. Zannetti,
J. Stat. Mech. P02013 (2008).
 
\bibitem{a-d3}
M. Pleimling, A. Gambassi,
Phys. Rev. B {\bf 71}, 180401(R) (2005).

\bibitem{Xinfty-d2}
P. Mayer, L. Berthier, J. P. Garrahan and P. Sollich, Phys. Rev. E
{\bf 68}, 016116 (2003).
C. Chatelain, J . Phys. A {\bf 36}, 10739 (2003).
F. Sastre, I. Dornic and H. Chat\'e, Phys. Rev. Lett. {\bf 91}, 267205 (2003).
C. Chatelain, J. Stat. Mech. P06006 (2006).

\bibitem{split}
L. F. Cugliandolo, {\it Dynamics of glassy systems}, in 
Les Houches Session 77, arXiv:cond-mat/0210312.

\bibitem{aging-sub-critical}
L. F. Cugliandolo and D. S. Dean, J. Phys. A {\bf 28}, 4213  (1995);
J. Phys. A {\bf 28}, L453 (1995). 
L. F. Cugliandolo, J. Kurchan, and G. Parisi, J. Phys. (France) {\bf 4}, 1641 (1994).
A. Barrat, Phys. Rev. E {\bf 57}, 3629 (1998).
L. Berthier, J-L Barrat, and J. Kurchan, Eur. Phys. J. B {\bf 11}, 635 (1999). 

\bibitem{nota}
For Ising spins. For vector spins or soft spins (Langevin equation)
$\chi _{st}$ does not vanish.

\bibitem{Xinfty-d1} 
E. Lippiello and M. Zannetti, Phys. Rev. E {\bf 61}, 3369 (2000).
C. Godr\`eche and J.-M. Luck, J. Phys. A {\bf 33}, 1151 (2000).

\bibitem{chatelain-ricci} C. Chatelain, J. Phys. A {\bf 36}, 10739  (2003).
F. Ricci-Tersenghi, Phys. Rev. E {\bf 68}, 065104(R) (2003). 
L. Berthier, Phys. Rev. Lett. {\bf 98}, 220601 (2007).

\bibitem{algo}
E. Lippiello, F. Corberi, M. Zannetti, Phys. Rev. E {\bf 71}, 036104 (2005). 

\bibitem{Aron}
C. Aron, C. Chamon, L. F. Cugliandolo and M. Picco,
J. Stat. Mech. (2008) P05016. 

\bibitem{Mayer} 
P. Mayer, H. Bissig, L. Berthier, L. Cipelletti, J. P. Garrahan, P. Sollich, and V. Trappe, 
Phys. Rev. Lett. {\bf 93}, 05002 (2005). P. Mayer, P. Sollich, L. Berthier, and 
J. P. Garrahan, J. Stat. Mech. P05002 (2005). 

\bibitem{ontheconnection}
E. Lippiello, F. Corberi, M. Zannetti, Eur. Phys. J. B {\bf 24}, 359 (2001). 

\bibitem{NBF}
F.~Corberi, E.~Lippiello, and M.~Zannetti, Phys. Rev. E {\bf 63}, 061506 (2001);
F. Corberi, E. Lippiello and M. Zannetti, Phys. Rev. E {\bf 72}, 056103 (2005);
F.~Corberi, E.~Lippiello, and M.~Zannetti, Phys. Rev. E {\bf 74}, 041113 (2006);
F.~Corberi, E.~Lippiello, and M.~Zannetti, Phys. Rev. E {\bf 74}, 041106 (2006);
R.~Burioni, D.~Cassi, F.~Corberi, and A.~Vezzani, Phys. Rev. Lett. {\bf 96}, 235701 (2006);
R.~Burioni, D.~Cassi, F.~Corberi, and A. Vezzani, Phys. Rev. E {\bf 75}, 011113 (2007).

\bibitem{Castillo}
A. Parsaeian and H. E. Castillo,
{\it  Universal fluctuations in the relaxation of structural glasses},
 arXiv:0811.3190;
{\it Equilibrium and non-equilibrium fluctuations in a glass-forming liquid}
 arXiv:0802.2560;
 Phys. Rev. E 78, 060105(R) (2008).   
H. E. Castillo and A. Parsaeian, Nature Physics {\bf 3}, 26 (2007).

\bibitem{Joseluis} J. L. Iguain, S. Bustingorry and L. F. Cugliandolo,
in preparation.

\bibitem{Ciliberto} S. Joubaud, B. Percier, A. Petrosyan, and S. Ciliberto, 
arXiv:0810.1392.

\bibitem{temp0}
E. Lippiello, F. Corberi, M. Zannetti, Phys. Rev. E {\bf 78}, 011109 (2008). 



\end{thebibliography}
\end{document}